\documentclass[aps,prb,a4paper,twocolumn]{revtex4}
\usepackage{amsmath,epsfig}

\makeatletter
\def\@dotsep{4.5}
\makeatother

\renewcommand{\_}[1]{\overline{#1}}
\renewcommand{\=}[1]{\_{\_{#1}}}
\renewcommand{\.}{\cdot}
\renewcommand{\l}[1]{\label{eq:#1}}
\renewcommand{\r}[1]{(\ref{eq:#1})}
\renewcommand{\Re}{\rm Re}

\begin{document}

\title{Subwavelength microscope that uses frequency scanning for image reconstruction}
\author{Stanislav Maslovski}
\affiliation{St.~Petersburg State Polytechnical Univ., Radiophysics Dept., Polytechnicheskaya 29, 195251, St.~Petersburg, Russia}
\email[]{stanislav.maslovski@gmail.com}
\author{Pekka Alitalo, and Sergei Tretyakov}
\affiliation{Dept.\ of Radio Science and Engineering,
P.O.\ Box 3000,
FI-02015 TKK, Finland}
\email[]{sergei.tretyakov@tkk.fi}

\date{\today}

\begin{abstract}
A new principle of subwavelength imaging based on frequency scanning
is considered. It is shown that it is possible to reconstruct the
spatial profile of an external field exciting an array (or coupled
arrays) of subwavelength-sized resonant particles with a frequency
scan over the whole band of resonating array modes. During the scan
it is enough to measure and store the values of the near field at
one or at most two points. After the scan the distribution of the
near field can be reconstructed with simple post-processing. The
proposed near-field microscope has no moving parts.
\end{abstract}


\maketitle

\section{Introduction}

The maximum achievable resolution of any conventional lens, the
function of which is based on focusing of electromagnetic radiation,
is the wavelength of the used radiation (the limit that is also known as
the diffraction limit or the Rayleigh limit). The resolution limit
is based on the fact that the evanescent waves, which carry the
information about the subwavelength properties of the source, are
lost due to their exponential attenuation,  and only the propagating
waves are restored in the image plane by a focusing lens.
The resolution of an imaging device can be greatly
improved if this device is able to react to the evanescent part of
the spatial spectrum of the incident field. There are several known
techniques which allow subwavelength-resolution imaging: superlenses
based on the use of materials with negative parameters,\cite{Pendry,Fang}
superlenses based on phase-conjugation
(non-linear three-wave mixing),\cite{phase} arrays of small resonant
particles.\cite{grids,enlarging,pekka,line,Freire} However, the only mean to
actually measure the image-plane near-field distributions at subwavelength scale
is the scanning near-field microscope which uses a small moving
probe. Recently, hyperlenses which
transform evanescent modes into propagating ones have been
proposed\cite{narimanov,smol,hyper} with the goal to develop subwavelength
microscopes. These devices allow image detection by a
stationary system like a charge-coupled device (CCD) matrix, however,
the manufacturing technology for hyperlens structures needs
further development.

In this paper we introduce a near-field subwavelength imaging device
that is based on frequency scanning with following post-processing
of measured data. The field is measured only at one or two points in
space, and the role of sensor is played by an electrically dense grid of small resonant
particles. The device does not use any moving probes, allowing for
very fast subwavelength imaging. In this device the conventional measuring of near fields
at many points in space is replaced by measuring resonantly enhanced near fields
at many frequency points, which allows us to calculate the
spatial field distribution. The maximum spatial resolution of the
device is limited by the period of the sensing grid and depends on
the quality factor of grid particles as well on the number of
particles.

The present method is a development of the approach of our
previous paper\cite{grids} which  introduced a system of two coupled resonant
grids or arrays of small inclusions as a superlens capable for
resonant amplification of evanescent fields and creation of images
with subwavelength detail. That idea was further extended for
enlarging superlenses.\cite{enlarging} Possible realization in
the optical region was also demonstrated.\cite{pekka,line}
Experimental confirmations in the microwave region were made using
grids of small resonant electric dipoles\cite{grids,enlarging} and
magnetic dipoles (split rings).\cite{Freire}
Grids of resonant
particles get strongly excited when the transversal
wavenumber $k_t$ of an incident evanescent wave matches the
propagation factor of grid's surface wave. We showed both
theoretically and experimentally how this effect
can be used for resonant amplification of evanescent
fields. Because of the resonance, the amplitude of the evanescent
wave in the image plane of a dual-grid structure\cite{grids}
becomes equal to the amplitude of the evanescent incident wave in
the source plane.

\begin{figure}[h!]
\centering
\epsfig{file=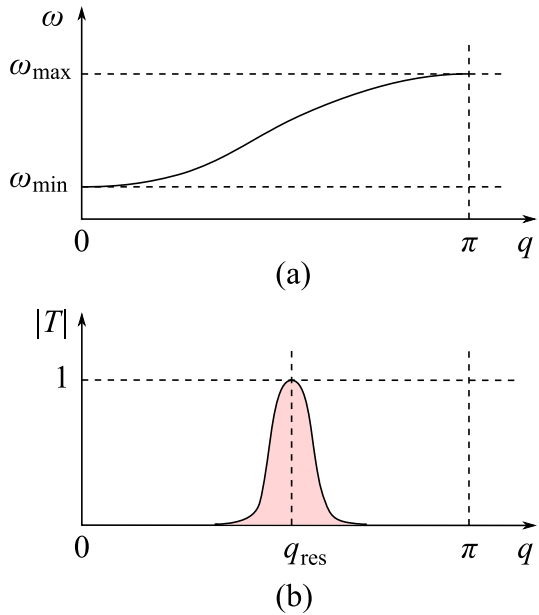,width=7cm}
\caption{a) An example dispersion curve of a periodic structure. On the horizontal
axis $q = k_td$. b)  Transmission trough a pair of resonating arrays
at a given frequency as a function of $q$.}
\label{figdisp}
\end{figure}

Matching of the propagation factors takes place at a certain
frequency. The grid of densely packed resonant particles can be
modeled by its grid impedance $Z_{\rm g}(\omega, k_{\rm t})$, which
connects the surface-averaged tangential electric and magnetic
fields.\cite{modeling} The grid impedance depends on the
frequency $\omega$ and the propagation factor along the grid plane
$k_{\rm t}$. The dispersion equation for the surface waves on an
impedance grid in free space can be written as
\begin{equation}
Z_{\rm g}(\omega, k_{\rm t}) + {Z_0(\omega, k_{\rm t})\over 2} = 0,
\l{disp}
\end{equation}
where $Z_0(\omega, k_t)$ is the wave impedance of the corresponding
free-space plane wave:
\begin{eqnarray}
Z_0(\omega, k_{\rm t})=\eta_0{\sqrt{k_0^2-k_t^2}\over k_0},\qquad
\text{TM modes},\\
Z_0(\omega, k_{\rm t})=\eta_0{k_0 \over \sqrt{k_0^2-k_t^2}},\qquad
\text{TE modes},
\end{eqnarray}
where $\eta_0=\sqrt{\mu_0\over \epsilon_0}$ and
$k_0=\omega\sqrt{\mu_0\epsilon_0}$. The solutions of this equation
are pairs $(\omega, k_t)$ that lie on grid's dispersion curve.  If
the grid is periodic with the period $d \ll \lambda_0$ ($\lambda_0$
is the free-space wavelength), its dispersion curve usually looks
like it is shown in Fig.~\ref{figdisp}(a).

We see that for an operating frequency $\omega \in [\omega_{\rm
min}, \omega_{\rm max}]$ there is only a single resonating value
for the propagation constant $k_t$.
Scanning the frequency from $\omega_{\rm min}$ up to $\omega_{\rm
max}$ we can sequentially excite the surface modes of all possible
$k_t$ ranging from 0 to ${k_t}_{\rm max} = \pi/d$. The amplitude of
an excited mode is proportional to the amplitude of the
corresponding spectral component of the incident field. Knowing the
spatial profile of every surface mode (from the theory or from an
initial calibration measurement) we can find the amplitude of the
mode from a field measurement done only at a single point in the
image plane (or at most at a couple of discrete points, to account
for degenerate cases of zero field of certain modes at the
measurement point).
In a finite-size grid the spectrum of surface modes is discrete.
In this case the field measurement is done at discrete frequencies.

It is worth noting that for open structures (in contrast to setups
confined inside a closed impenetrable cavity) the surface modes with
$0 \le k_t \le k_0 \ll \pi/d$ are leaky modes, i.e., they radiate
into surrounding space. However, the operation of the microscope is
based on the use of eigenmodes with high propagation constants
($k_t\gg k_0$), which are strongly bound to the grid surface and
radiate only at inhomogeneities and at the grid ends. Moreover, in
some resonant systems such as the metasolenoid\cite{metasol} even
the modes with small values of $k_t$ produce a strong resonant
response because of a high Q-factor of its resonances ($Q$ is of the
order of $10^3$).

In a system of two coupled resonant grids\cite{grids} the amplitude of the field in the image plane
can be made the same as in the source plane so that the system works as a ``perfect lens''\cite{Pendry}
at every resonant frequency of the system. This allows us to find the spatial
field distribution of the unknown source by weighted integration of measured values, as is explained in
Section~\ref{two_grids}.
In principle one  can use just a single resonating grid. The only
difficulty here is that the relation between the amplitude of the
excited surface mode and the unknown external field is more complex.
The single-grid microscope is explained in Section~\ref{one_grid}.

\section{Principle of spatial imaging by frequency scanning}
\label{two_grids}

\begin{figure}[h!]
\centering \epsfig{file=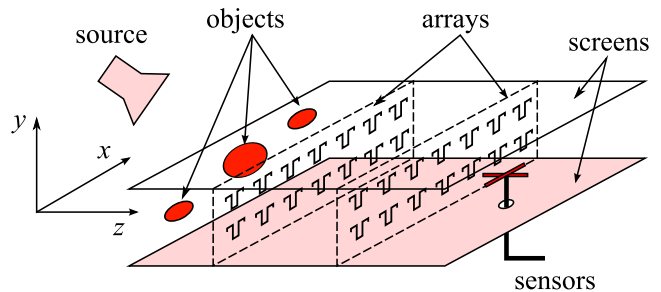, width=8.5cm} \caption{The
side view of a possible lens structure.  Two arrays of small
particles are positioned along the $x$-axis. The screening box is
open at one of its sides (at the source plane).}
\label{figgeom}
\end{figure}

For clarity, let us consider a lens composed of a couple  of arrays
of small  resonating electric dipoles (oriented along the $x$-axis)
placed in between two metal screens: the same system as we used in
our measurements.\cite{grids} The dispersion curves of the
arrays look like in Fig.~\ref{figdisp}(a) and the field is measured at
the the middle point of the image plane at $x = 0$. To cope with the
situation when the $x$-component of the electric field vanishes at
this point for a certain mode it is necessary to have a couple of
sensors there: one measures $E_x$ and the other measures the
orthogonal component $E_z$ (the longitudinal component). We will
also assume that the arrays consist of many particles, so that the
spectrum of its modes is practically continuous. An example geometry
of the structure is shown in Fig.~\ref{figgeom}.

Let us assume that the object under study is located in the source
plane and it can be excited at the frequencies in the range
$[\omega_{\rm min}, \omega_{\rm max}]$ and that the properties of
the source itself do not change much when we scan the frequency in
this range. Practically this means that we have to provide as narrow
frequency band of the dispersion curve as possible. In terms of the
design of the grids this leads to the known conditions:\cite{grids}
the particles must be weakly interacting and be high-Q
resonators themselves. At microwave frequencies, electrically small
copper-wire meanders are good enough with
enough narrow resonant band.\cite{grids} In the optical range
plasmonic metal spheres can be used.\cite{pekka,line}

If the source is passive (for example, it is a collection of small
pieces of metal) it can be excited with an antenna that illuminates
it with a plane wave of the necessary frequency. Because the
structures we deal with are most effectively excited with evanescent
waves, the illuminating source will not influence the operation of
the device (or at least its influence is predictable and can be
taken into account in the post-processing stage). It can be a good
engineering task to figure out the optimal disposition of the
illuminating source and the arrays. In the particular example
considered here (Fig.~\ref{figgeom}), it is assumed that the
source electric field is polarized along $x$ and traveling modes are not
supported inside the screening box.

For a  given frequency $\omega \in [\omega_{\rm min}, \omega_{\rm
max}]$ the transmission coefficient through the lens as a function
of $k_x$ will look like in Fig.~\ref{figdisp}(b). The characteristic
width of the spike is physically determined by the characteristic
size $L$ (the array length) of the lens: $\Delta k_x \approx \pi/L$.
In principle the spike can be made as narrow as required. The field
sensor will measure a kind of average amplitude of the modes with
${k_x}_{\rm res} - \Delta k_x/2 < |k_x| < {k_x}_{\rm res} + \Delta
k_x/2$.

The device operates  as follows. We scan the frequencies from
$\omega_{\rm min}$ up to $\omega_{\rm max}$ and store the measured
electric field complex amplitudes at every frequency. In the
post-processing we use the known spatial profiles of the modes to
restore the actual field distribution from the measured field values
at the source plane. This is possible via simple integration,
because the amplitudes of the fields of every mode are the same at
the image and source planes (neglecting dissipation in the
grids).\cite{grids} Denoting for even modes: $A^{\rm even} = E_x^{\rm
probe}$,  and for odd modes: $A^{\rm odd} = - (k_z(\omega) /
k_x(\omega)) E_z^{\rm probe}$, the field distribution in the source
plane can be expressed as a sum of all modes:

\begin{multline}
E_x(x) = {L\over\pi} \int\limits_{\omega_{\rm min}}^{\omega_{\rm
max}}\Big[ A^{\rm even}(\omega)\cos(k_x(\omega)x)\\
+ A^{\rm odd}(\omega)\sin(k_x(\omega)x) \Big]{d k_x(\omega) \over
d\omega}\, d\omega.
\end{multline}
The coefficient $L/\pi$ accounts for the spectral density of modes.

As it was said above, in a finite-size grid the spectrum of surface
modes is discrete.  In this case the field measurement is better to
be done at discrete frequencies. We will consider such a system in
the next section.

\section{Single array of resonant particles as a near-field sensor}
\label{one_grid}

\begin{figure}[h!]
\centering \epsfig{file=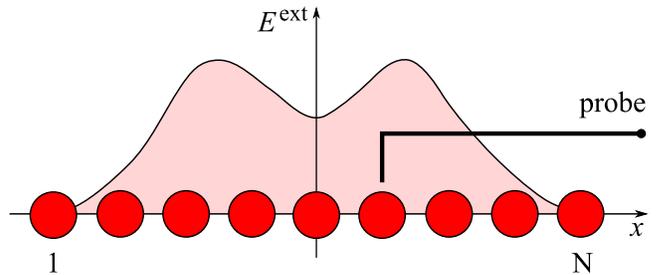, width=8.5cm}
\caption{A linear array of plasmonic nanoparticles as a
sub-wavelength microscope sensor. The line shows the electric-field
profile to be determined by the microscope. The probe is
stationary.} \label{one}
\end{figure}

Let us consider a linear array of resonating particles,
Fig.~\ref{one}. The particles are electrically small passive
inclusions that we model by electric dipoles with known
polarizabilities. We assume that there are $N$ particles in the
array and that the array period is $d$. The electric dipole moments
$p_k$ of the array particles can be found from the following system
of linear equations:
\begin{equation}
\alpha^{-1}(\omega)\,p_k - \sum_{n=1,\ n\neq k}^{N}\beta_{n-k}(\omega)p_n = E^{\rm ext}_k,
\l{system}
\end{equation}
where $E^{\rm ext}_k$ is the external electric field at the position
of the particle number $k$, $\alpha(\omega)$ is the dipole
polarizability, and $\beta_{n-k}(\omega)$ accounts for interaction
in a pair of dipoles with indices $n$ and $k$. Our goal is to find
the spatial distribution of the external field (that is, find vector
$E^{\rm ext}_k$) from measured field at one point but at several
frequencies.

For dipoles orthogonal to the line of the array
\begin{equation}
\beta_{m}(\omega) = {1\over 4\pi\varepsilon_0}\left[{k_0^2\over |m|d} - {jk_0\over |m|^2d^2} - {1\over |m|^3d^3}\right]
e^{-jk_0|m|d}.
\end{equation}
For dipoles parallel to the line of the array
\begin{equation}
\beta_{m}(\omega) = {1\over 2\pi\varepsilon_0}\left[{jk_0\over |m|^2d^2} + {1\over |m|^3d^3}\right]
e^{-jk_0|m|d}.
\end{equation}
The polarizability of a small resonant dipole can be approximated as
\begin{equation}
{1\over \alpha(\omega)} = {1\over \alpha_0}\left[{\omega_0^2-\omega^2\over \omega_0^2} + {j\omega\over \omega_0}{1\over Q} \right] + {j\eta_0\omega^3\over 6\pi c^2},
\end{equation}
where $\alpha_0$ is the  static polarizability, $\omega_0$ is the
resonant frequency, $Q$ is the quality factor of the particle and
the last term accounts for the radiation loss.\cite{modeling}

The system of equations \r{system} can be written in matrix form
\begin{equation}
\=A(\omega)\.\_p(\omega) = \_E^{\rm ext},
\end{equation}
where $\=A(\omega)$ is the system matrix, the elements of which can
be calculated from the above formulas; $\_p(\omega)$ and $\_E^{\rm
ext}$ are vectors of length $N$. The eigenmodes of the array can be
found from the equation
\begin{equation}
\det \=A(\omega) = 0.
\l{determinant}
\end{equation}
There are $N$ modes in total. In a system with non-zero loss the
modal frequencies  defined in this manner are complex quantities
with the imaginary part representing the decay of oscillations. In
our problem, however, the frequency is defined by the applied
external field and is real. Therefore, the array resonances are
found as the solutions of equation
\begin{equation}
\Re\{\det \=A(\omega_k)\} = 0,
\l{realdet}
\end{equation}
where all $\omega_k$ are real. Note that the matrix $\=A(\omega)$ is almost singular at the frequencies defined by \r{realdet}. Therefore, its inverse can be approximated as
\begin{equation}
\=A^{-1}(\omega_k) \approx \lambda^{-1}(\omega_k) \={\cal P}(\omega_k),
\end{equation}
where $\lambda(\omega_k)$ is the eigenvalue of $\=A(\omega)$ with the smallest magnitude at $\omega \rightarrow \omega_k$ and $\={\cal P}(\omega_k)$ is the projection operator for the corresponding eigenvector. Hence, the dipole moments at the resonances are
\begin{equation}
\_p(\omega_k) \approx \lambda^{-1}(\omega_k) \={\cal P}(\omega_k)\.\_E^{\rm ext}.
\end{equation}
From here we see that at the resonances the  array acts effectively
as a filter extracting a single spatial harmonic from the external
field.

At the resonances the induced dipole moments are maximal  and so is
the electric field at the probes that we place near the array.
Therefore, it makes sense to measure these fields at resonances. At
these frequencies $\omega_k$ the field at the location of a probe
can be found as
\begin{equation}
E(\omega_k) = \sum_{n=1}^{N}c_n(\omega_k) p_n(\omega_k),
\l{probe}
\end{equation}
where $c_n(\omega_k)$ are the coefficients describing coupling of the probe to each dipole in the array.

We can rewrite \r{probe} in matrix form
\begin{equation}
E(\omega_k) = \_c^{\rm T}(\omega_k)\.\_p(\omega_k) = \_c^{\rm T}(\omega_k)\.\=A^{-1}(\omega_k)\.\_E^{\rm ext},
\end{equation}
or, for all $N$ frequencies at once
\begin{equation}
\begin{pmatrix}
E(\omega_1)\\
\cdots\\
E(\omega_n)
\end{pmatrix} =
\begin{pmatrix}
\_c^{\rm T}(\omega_1)\.\=A^{-1}(\omega_1)\\
\cdots\\
\_c^{\rm T}(\omega_n)\.\=A^{-1}(\omega_n)
\end{pmatrix} \.
\begin{pmatrix}
E^{\rm ext}_1\\
\cdots\\
E^{\rm ext}_N
\end{pmatrix},
\l{fin_system}
\end{equation}
where the rows in the central matrix are products of $\_c^{\rm T}$
and $\=A^{-1}$ at  all the resonant frequencies. Let us recall that
we assume that the distribution of the external field does not
change when we scan the frequencies in the range~$[\omega_1,
\omega_N]$.

Finally we can solve \r{fin_system} for the distribution of the external field:
\begin{equation}
\begin{pmatrix}
E^{\rm ext}_1\\
\cdots\\
E^{\rm ext}_N
\end{pmatrix} =
\begin{pmatrix}
\_c^{\rm T}(\omega_1)\.\=A^{-1}(\omega_1)\\
\cdots\\
\_c^{\rm T}(\omega_n)\.\=A^{-1}(\omega_n)
\end{pmatrix}^{-1}\.
\begin{pmatrix}
E(\omega_1)\\
\cdots\\
E(\omega_n)
\end{pmatrix}.
\end{equation}
For this solution to exist the matrix in~\r{fin_system} must be
non-singular.  This can be achieved either by placing a probe at a
non-symmetric location with respect to the dipole array or by using
more than one probe and intermixing their responses
in~\r{fin_system}.

\section{Numerical example}

As an example let us calculate the response  of a linear array of
orthogonal dipoles and then restore the distribution of the external
field using the method introduced above. We choose the following
parameters for the dipole array: the period is such that $k_0d =
0.1$ at the dipole resonance, the number of dipoles in the array $N
= 21$, the quality factor of the particle itself (without radiation
loss which is accounted for separately) is $Q = 10^3$, the static
dipole polarizability is such that
$\alpha_0/(\varepsilon_0d^3)=0.5$. This value of the polarizability
is rather high. For a metal sphere it corresponds to the radius of
$a = d\sqrt[3]{0.5/(4\pi)}\approx 0.34\,d$. Note, however, that
resonant particles of complex shape or plasmonic particles can
provide high values of polarizability even if their size is electrically
quite small.

\begin{figure}[h!]
\centering \epsfig{file=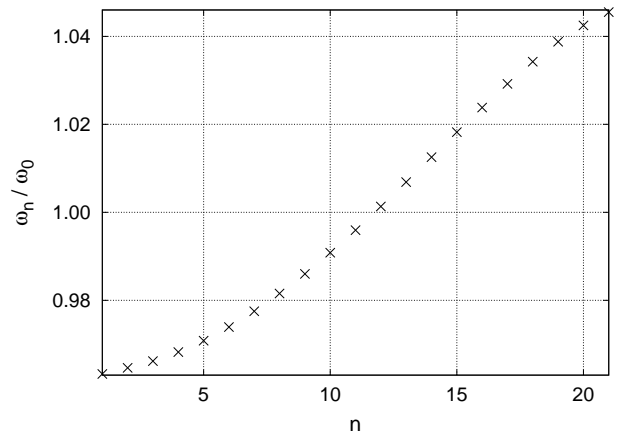,width=8.5cm}
\caption{Resonant frequencies of the dipole array.} \label{resfreq}
\end{figure}

In Fig.~\ref{resfreq} the resonant frequencies of the array are
plotted as function of their index varying from 1 to 21.  These
frequencies are found from \r{realdet}.

\begin{figure}[h!]
\centering \epsfig{file=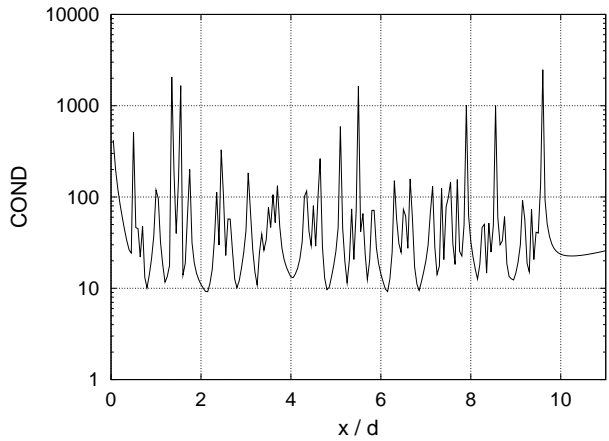,width=8.5cm}
\caption{Condition number of the system matrix as a function of the
probe location. Coordinate $x$ is along the array with the origin at
the middle dipole.} \label{figcond}
\end{figure}

The probe is  located close to the array at the distance of half a
period from the array line. For the probe to be able to react to
both symmetric and antisymmetric modes the probe must not be placed
in a symmetric position with respect to the array. To find the
optimal position we investigate how the condition number of the
matrix \r{fin_system} changes when we move the probe along the
array. This is shown in Fig.~\ref{figcond}.

The optimal position  is found to be in front of the second dipole
if counted from the middle of the array. At this point the condition
number is about~10. This means that the measurement errors or any
other noise of relative amplitude of $0.1$ will destroy the
restoration process. For practical purposes the measurement error
must not exceed $10^{-3}$--$10^{-2}$.

The higher is $Q$ the  lower is the condition number. The same holds
for the value of $\alpha_0$. From the other hand, the condition
number grows with increasing the number of dipoles in the array. For
an array of about 50 particles to obtain the condition number of 20
one has to provide $Q\sim 10000$ which is hardly achievable.
Therefore, with large arrays it would be more practical to employ
several probes placed at carefully selected positions.

\begin{figure}[h!]
\centering \epsfig{file=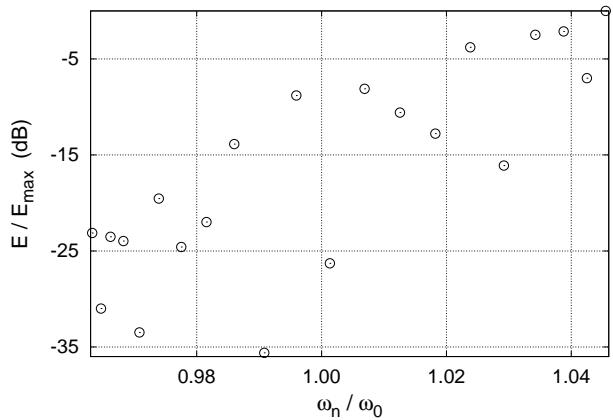,width=8.5cm}
\caption{Electric field amplitude at the location of the probe at
different resonant frequencies.} \label{figprobe}
\end{figure}

\begin{figure}[h!]
\centering \epsfig{file=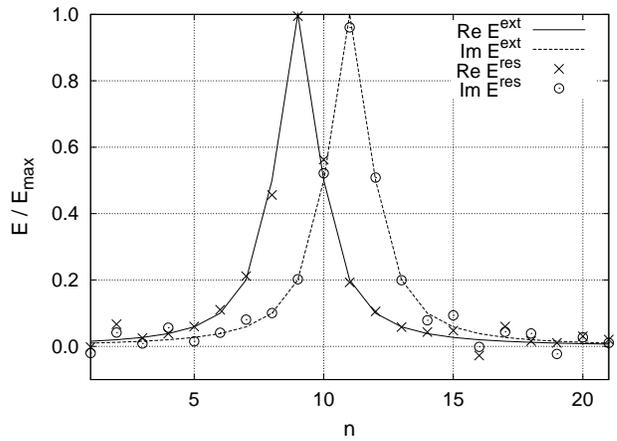,width=8.5cm}
\caption{The profiles of the external field (solid lines) and the restored field
(symbols).} \label{figrestored}
\end{figure}

In the following field restoration example we assume the following
distribution of the external field:
\begin{equation}
E^{\rm ext}_k = {1\over 1 + (k - 9)^2} + j{1\over 1 + (k - 11)^2}.
\l{profile}
\end{equation}
With such an external field in action  the amplitude of the electric
field at the location of the probe is shown in Fig.~\ref{figprobe}.
The probe measures the field only at the array resonant frequencies.
In Fig.~\ref{figrestored} the restored spatial profile of the
external field is compared with the original one given by
\r{profile}.  To show the effect of noise we have added a normally
distributed noise of the relative variance of $10^{-2}$ to the
measured probe field. If there is no noise, the restoration is
exact, that is, the values of the incident field at the positions of
each particle in the array are determined exactly. This means that
the microscope resolves spatial details down to the scale of one
array period.

\section{Conclusions}

In this paper we have presented a near-field imaging principle which
allows us to determine spatial distributions of electric or magnetic
field with sub-wavelength resolution using observations at several
frequencies within a narrow frequency range of sensor's
resonances. The field is measured only at one or a few selected
points behind the sensing array of resonant particles. The limit for
the image resolution is determined by the grid period. We have shown
that for reliable measurements it is desirable to use resonant
particles with high quality factors and high polarizability values.
The evanescent fields emanating from the object under study are
effectively enhanced due to resonant excitation of the sensor
particles. The necessary range of the frequency scan is reduced if
the array particles are weakly coupled to each other. Since the
device has no moving parts and there are known means to change the
frequency of the probing electromagnetic wave very quickly, this
method offers a possibility to make observations of complex
sub-wavelength objects very fast and in a non-invasive way. For
microwave applications, the sensor can be formed by electrically
small metal particles, e.g., in form of a meander\cite{grids} or as
dense packages of split rings (metasolenoids).\cite{metasol} For
optical applications, the sensor can be potentially realized as
grids of plasmonic nanoparticles.\cite{pekka,line} Another
possibility is to use grids of resonant nanocavities in thin metal
layers.\cite{holes} The last case has the advantage of reduced
coupling between resonant voids due to field decay in the metal
layer. We are now studying dispersion properties of arrays of
nanovoids in metal near an interface with free space with some
promising results.






\begin{thebibliography}{99}

\bibitem{Pendry}
J.~B.~Pendry, Phys.\ Rev.\ Lett.\ {\bf 85}, 3966 (2000).

\bibitem{Fang}
N.~Fang, H.~Lee, C.~Sun, and X.~Zhang, Science {\bf 308}, 534 (2005).

\bibitem{phase}
S.~Maslovski and S.~Tretyakov, J.\ Appl.\ Phys.\ {\bf 94}, 4241 (2003).

\bibitem{narimanov}
Z.~Jacob, L.~V.~Alekseyev, and E.~Narimanov,
Opt.\ Express {\bf 14}, 8247 (2006).

\bibitem{smol}
I.~I.~Smolyaninov, Y.-J.~Hung, and C.~C.~Davis, Science {\bf 315}, 1699 (2007).

\bibitem{hyper}
Z.~Liu, H.~Lee, Y.~Xiong, C.~Sun, and X.~Zhang, Science {\bf 315}, 1686 (2007).

\bibitem{grids}
S.~Maslovski, S.~Tretyakov, and P. Alitalo, J.\ Appl.\ Phys.\ {\bf 96}, 1293 (2004).

\bibitem{enlarging}
P.~Alitalo, S.~Maslovski, and S.~Tretyakov, Phys.\ Lett.\ A {\bf 357}, 397 (2006).

\bibitem{pekka}
P.~Alitalo, C.~Simovski, A.~Viitanen, and S.~Tretyakov, Phys.\ Rev.\ B {\bf 74}, 235425 (2006).

\bibitem{line}
C.~Simovski, A.~Viitanen, and S.~Tretyakov, J.\ Appl.\ Phys.\ {\bf 101}, 123102 (2007).

\bibitem{Freire}
M.~J.~Freire and R.~Marques, Appl.\ Phys.\ Lett.\ {\bf 86}, 182505 (2005).

\bibitem{modeling}
S.~Tretyakov, {\it Analytical modeling in Applied Electromagnetics} (Artech House, Boston, 2003).

\bibitem{metasol}
L.~Jylha, S.~Maslovski, and S.~Tretyakov, J.\ Electromagnet.\ Waves Appl.\ {\bf 19}, 1327 (2005).

\bibitem{holes}
S.~Coyle, M.~C. Netti, J.~J.~Baumberg, M.~A.~Ghanem, P.~R.~Birkin, P.~N.~Bartlett, and D.~M.~Whittaker, Phys.\ Rev.\ Lett.\ {\bf 87}, 176801 (2001).
\end{thebibliography}
\end{document}